\documentclass[aps,pra,epsfigure,twocolumn,showpacs]{revtex4}
\usepackage{amsmath}
\usepackage{amstext}
\usepackage{latexsym}
\usepackage{psfig}
\usepackage{graphicx}
\usepackage{amsfonts}

\newcommand{\ket}[1]{\left\vert#1\right\rangle}
\newcommand{\modul}[1]{\left\vert#1\right\vert}

\newcommand{\one}{\mbox{$1 \hspace{-1.0mm}  {\bf l}$}}

\newcommand{\pro}[2]{\left\vert#1\rangle\langle#2\right\vert}

\newcommand{\bra}[1]{\left\langle#1\right\vert}
\newcommand{\sand}[3]{\left\langle#1\vert#2\vert#3\right\rangle}

\begin{document}

\title{On the transfer of entanglement from a two-mode squeezed state to a pair of qubits}
\author{M. Paternostro, W. Son and M. S. Kim}
\affiliation{School of Mathematics and Physics, The Queen's University,
Belfast BT7 1NN, United Kingdom}
\date{\today}

\begin{abstract}
There have recently been interests in transferring entanglement between two quantum systems in different Hilbert spaces. In particular, the study of entanglement transfer from a continuous-variable to a qubit system has a primary importance due to practical implications. A continuous-variable system easily propagates entanglement while a qubit system is easy to manipulate. We investigate conditions to entangle two two-level atoms using a broad-band two-mode squeezed field driving the cavities where the atoms are.
\end{abstract}
\maketitle


{\it Introduction - } The possibility to exploit the correlations transferred from a two-mode squeezed state of light to two distant atoms, initially prepared in a separable state, appeals for the purposes of quantum communication and distributed quantum computation~\cite{wonmin}. On the other hand, some of the aspects that characterize the dynamics of two-level atoms in a squeezed reservoir have been studied and exploited for quantum state engineering~\cite{massimoinfanzia}. Very recently, Kraus and Cirac~\cite{kraus} have suggested a cavity-quantum-elctrodynamics (CQED) system in which two trapped atoms in respective optical cavities interact with an external broad-band two-mode squeezed state. The scheme has been exploited to produce an entangled atomic state.

Starting from this system,  we address the conditions that have to be satisfied to perform effective entanglement-transfer from a continuous-variable system to a pair of qubits. This process is intriguing because it combines the manipulability of a qubit system to the possibility of entanglement propagation via a correlated state of light. We analyze the entangling properties of the master equation that describes the dynamics of the system in~\cite{kraus}. We find bounds on the degree of purity and entanglement of the two-mode driving field, assessing the corresponding effects on the reduced state of the two atoms, both for dynamical and steady conditions. We consider the cavity decay rate and the atomic spontaneous emission for the efficiency of the scheme. 

{\it Master equation - }Here, we briefly review the model recently proposed in~\cite{kraus}. Two identical two-level atoms are respectively trapped in spatially separated cavities, which are externally driven by a broad-band two-mode squeezed state of light characterized by the phase-insensitive factor $N$ and the phase-sensitive one $M$. The two identical cavities have an energy decay rate $\kappa$. Under the assumption that the bandwidth $\Delta\omega_{ext}$ of the squeezed field is greater than the cavity decay rate $\kappa$, the Born-Markov approximation is applied. The master equation reads ($\hbar=1$)
\begin{equation}
\label{master1}
\partial_{t}\rho=-i[\hat{H}_{a1}+\hat{H}_{b2},\rho]+{\cal \hat{L}}_{cav}\rho,
\end{equation} 
with $\hat{H}_{a1}=\Omega\left(\hat{\sigma}^{+}_{1}\hat{a}+\hat{a}^{\dag}\hat{\sigma}^{-}_{1}\right)$ a Jaynes-Cummings interaction Hamiltonian between atom $1$ and cavity mode $a$ (analogously for $\hat{H}_{b2}$). Here, $\hat{\sigma}^{+}_{1}=(\hat{\sigma}^{-}_{1})^{\dag}=\ket{e}_{1}\!\bra{g}$ and $\Omega$ is the Rabi frequency. The Liouville operator ${\cal \hat{L}}_{cav}$ describes the interaction of the cavity-mode fields with the external broad-band squeezed field as 
\begin{equation}
\label{super1}
\begin{aligned}
\hat{\cal L}_{cav}\rho&=\kappa\!\sum^{b}_{\alpha=a}\left\{(N+1)(\hat{\alpha}\rho\hat{\alpha}^{\dag}-\hat{\alpha}^{\dag}\hat{\alpha}\rho)+N(\hat{\alpha}^{\dag}\rho\hat{\alpha}-\hat{\alpha}\hat{\alpha}^{\dag}\rho)\right\}\\
&+2\kappa{M}(\hat{a}\rho\hat{b}+\hat{b}\rho\hat{a}-\hat{b}\hat{a}\rho-\rho{\hat{a}^{\dag}{\hat b}^{\dag}})+h.c.,
\end{aligned}
\end{equation}
where $h.\,c.$ denotes hermitian conjugate of all the terms and $\rho$ contains the degrees of freedom of both the atoms and the cavity fields. The contributions from the spontaneous emission of the atoms from the excited state $\ket{e}_{a,b}$ have been neglected. However, they will be included later. 

For the case of a minimum-uncertainty squeezed state with a real squeezing parameter $r$, it is $N=\sinh^2{r},\, M=\sinh{r}\cosh{r}$ and  $M^2=N(N+1)$. Using the squeezing operator $\hat{S}(r)=\exp\left\{r\left(\hat{a}^{\dag}\hat{b}^{\dag}-\hat{a}\hat{b}\right)\right\}$ (assuming $r$ real)~\cite{knight} it is possible to transform $\rho$ as $\tilde{\rho}=\hat{S}(r)\rho\hat{S}^{\dag}(r)$, then Eq.~(\ref{master1}) becomes $\partial_{t}\tilde{\rho}=\left({\cal \hat{L}}_{a}+\hat{\cal L}_{cav,0}\right)\tilde\rho$.
Here we have defined ${\cal \hat{L}}_{a}\tilde\rho=-i[\hat{{\cal H}}_{a1}+\hat{\cal{H}}_{b2},\tilde\rho]$, where $\hat{\cal H}_{a1}=\hat{S}(r)\hat{H}_{a1}\hat{S}^{\dag}(r)$ (analogously for $\hat{{\cal H}}_{b2}$) and, moreover, $\hat{\cal L}_{cav,0}:=\hat{\cal L}_{cav}$ when $N=M=0$.

{\it Trap state - }It is straightforward to see that the Hamiltonian $\hat{\cal H}_{a1}+\hat{\cal H}_{b2}$ has the eigenstate $\ket{\Psi}=\ket{\psi}_{12}\otimes\ket{00}_{ab}$, where
$\ket{\psi}_{12}=\left(\sqrt{\frac{N+1}{2N+1}}\ket{gg}+\sqrt{\frac{N}{2N+1}}\ket{ee}\right)_{12}$.
From the Liouvillian $\hat{\cal L}_{cav,0}$ it is possible to identify the {\it jump operators} $\hat{C}_{1}=\sqrt{\kappa}\hat{{a}},\, \hat{C}_{2}=\sqrt{\kappa}\hat{{b}}$ that are responsible for the open dynamics of the cavity modes. Obviously, $Tr\left(\sum^{2}_{i=1}\hat{C}_{i}\pro{\Psi}{\Psi}\hat{C}^{\dag}_{i}\right)=0$ as $\hat{C}_{i}\ket{\Psi}=0$, so that $\ket{\Psi}$ is a steady state solution with the jump rate equal to zero. Hence, we call $\ket{\Psi}$ a trap state of the system. For a closed dynamics ($\kappa=0$), the only way to populate the trap state is to initially prepare the system in the trap state itself. In the limit of $\kappa\ll\Omega$, on the other hand, the probability that the system, after its transient period, will end up in the trap state $\ket{\Psi}$ is not negligible only if the system is actually prepared in this state. In general, we find that the population of the trap state changes according to 
$\partial_{t}(\sand{\Psi}{\rho}{\Psi})=\sum_{i=1,2}\sand{\Psi}{\hat{C}_{i}\rho\hat{C}^{\dag}_{i}}{\Psi}\propto\kappa$.
The rate at which the trapping state is populated because of the open dynamics of the system  is the decay rate $\kappa$. Indeed, $\kappa$ is the rate at which the quantum correlations between the cavities can be built up. It is, thus, convenient to choose a value $\kappa\gg\Omega$ to increase the probability that the atomic steady state is populated. This condition corresponds to a {\it bad-cavity regime}. However, $\kappa$ can not be increased at will because the relation $\Delta\omega_{ext}\gg\kappa$ represents, in general, a strong constraint for the validity of Eq.~(\ref{master1}). Typically, in fact, it is $\Delta\omega_{ext}\simeq\kappa/6\simeq2\pi\times12\, MHz$, in experiments in which squeezed light is coupled to a cavity-atom system~\cite{turchettekimble}. However, an analysis in the finite-bandwidth case is out of the scope of the current work and the condition $\Delta\omega_{ext}\gg\kappa$ will be assumed throughout this work. 

{\it Bad-cavity limit - } We now assume the bad-cavity limit $\kappa\gg\Omega,\Gamma$, where $\Gamma$ is the atomic spontaneous emission rate. Values $(\Omega,\Gamma)/2\pi\simeq(20,\,3.5)\, MHz$ allow for the squeezed state to build up inside the cavity~\cite{turchettekimble}. In this {\it{weak-coupling regime}}, the dynamics of the cavity modes interacting with the external driving fields is much faster than their interaction with the atoms. In these conditions, the atoms see the cavity modes in a steady state that is not affected by the atom-cavity dynamics. Thus, for the interaction time such that $\tau=\kappa{t}\gg{1}$, we take the steady state of the cavity modes to be a two-mode squeezed state as in the case without atoms being in the cavities. 
We are interested in the atomic evolution so to eliminate the cavity modes. In virtue of the above consideration, the adiabatic elimination of the cavity fields is possible by defining the {\it{projection}} operator~\cite{gardinerstochastic} as ${\cal P}\tilde\rho:~\tilde\rho\longrightarrow{\cal P}\tilde\rho=\rho_{ss}\otimes{Tr}_{ab}(\tilde\rho)=\rho_{ss}\otimes\rho_{12}$.
Here $\rho_{12}$ is the atomic density operator. It is easy to prove that ${\cal P}\hat{\cal L}_{a}{\cal P}\tilde\rho=0$ and it leaves the master equation in the adiabatically eliminated form $\partial_{t}\rho_{12}={Tr}_{ab}\left\{\hat{\cal L}_{a}\int^{\infty}_{0}e^{\hat{\cal L}_{cav,0}t}\hat{\cal L}_{a}\left(\rho_{ss}\otimes\rho_{12}\right)dt\right\}$.
To further reduce this equation we have to invoke the first Born-Markov approximation and exploit the relation $\hat{\cal L}_{cav,0}(\rho_{ss})=0$.
Going back to the initial unsqueezed picture, the adiabatically eliminated master equation takes the form
\begin{equation}
\label{master6}
\begin{aligned}
\partial_{t}\rho_{12}&=\frac{\gamma}{2}\sum^{2}_{i=1}\left[(N+1)\left(\hat{\sigma}^{-}_{{i}}\rho_{12}\hat{\sigma}^{+}_{{i}}-\hat{\sigma}^{+}_{{i}}\hat{\sigma}^{-}_{{i}}\rho_{12}\right)\right.\\
&+N\left(\hat{\sigma}^{+}_{{i}}\rho_{12}\hat{\sigma}^{-}_{{i}}-\hat{\sigma}^{-}_{{i}}\hat{\sigma}^{+}_{{i}}\rho_{12}\right)\left.\right]+\gamma{M}\left(\hat{\sigma}^{-}_{{1}}\rho_{12}\hat{\sigma}^{-}_{{2}}\right.\\
&\left.+\hat{\sigma}^{-}_{{2}}\rho_{12}\hat{\sigma}^{-}_{{1}}-\hat{\sigma}^{-}_{{2}}\hat{\sigma}^{-}_{{1}}\rho_{12}-\rho_{12}\hat{\sigma}^{+}_{{1}}\hat{\sigma}^{+}_{{2}}\right)+h.\,c.
\end{aligned}
\end{equation}
We have introduced the effective energy decay rate $\gamma=2\Omega^2/\kappa$. This master equation is in agreement with the result found in~\cite{kraus}. Eq.~(\ref{master6}) describes the effective decay of a pair of two-level atoms in a squeezed vacuum~\cite{massimoinfanzia}. The same equation can be obtained relaxing the condition of a minimum-uncertainty two-mode squeezed driving field, even if the calculations involved are more complicated. 

In this analysis we can include the Liouvillian terms related to the scattering of the atoms. These terms depend exclusively on the atomic operators according to
$\hat{\cal L}_{spont,i}\rho={\Gamma}/{2}\left(\hat{\sigma}^{-}_{i}\rho\hat{\sigma}^{+}_{i}-\hat{\sigma}^{-}_{i}\hat{\sigma}^{+}_{i}\rho-\rho\hat{\sigma}^{+}_{i}\hat{\sigma}^{-}_{i}+h.c.\right)$,
with $i=1,2$. The application of the projector operator ${\cal P}$ to these terms of the master equation simply gives $\rho_{ss}\otimes\hat{\cal L}_{spont,i}(\rho_{12})$. Following the lines sketched above, we get a generalized master equation of the same form as Eq.~(\ref{master6}) but with the parameters $\gamma'=\Omega^2(2+\Gamma)/\kappa$, $N'=CN/(1+C)$ and $M'=CM/(1+C)$ instead of $\gamma,\,N$ and $M$, respectively. Here we have introduced the {\it cooperativity parameter} $C=2\Omega^2/\Gamma\kappa$~\cite{turchettekimble}. With these new parameters it is always $M'<\sqrt{N'(N'+1)}$, even for a minimum-uncertainty two-mode driving field~\cite{massimoinfanzia}. Thus, inside the cavity a pure two-mode squeezed state can never be obtained, in a realistic situation. 

{\it Bound conditions for entanglement-transfer - }It is widely believed that an interaction mediated by a non-Markovian reservoir can set entanglement between two subsystems because of the memory effects of the bath. However, recently it was shown that even a Markovian reduced dynamics is able to set quantum correlation between two initially non interacting systems~\cite{benatti}. Under the hypothesis of weak-coupling and Markovian nature, the reduced dynamics of two systems, $A$ and $B$, interacting with a common bath can be cast into a master equation form as $\partial_{t}\rho_{AB}=-i[\hat{H}_{total},\rho_{AB}]+\hat{\cal L}\rho_{AB}$, with 
\begin{equation}
\label{benatti}
\hat{\cal L}\rho_{AB}=\sum^{6}_{\alpha,\beta=1}D_{\alpha\beta}\left(\hat{{\cal O}}_{\alpha}\rho_{AB}\hat{{\cal O}}_{\beta}-\frac{1}{2}\left\{\hat{{\cal O}}_{\beta}\hat{{\cal O}}_{\alpha},\rho_{AB}\right\}\right).
\end{equation}
Here, $\hat{{\cal O}}_{\alpha}=\sigma_{\alpha}\otimes\one$ for $\alpha=1,2,3$ and $\hat{{\cal O}}_{\alpha}=\one\otimes\sigma_{\alpha-3}$ for $\alpha=4,5,6$, with $\sigma_{1,2,3}$ the Pauli matrices and {\bf D} a $6\times6$ matrix decomposed as
\begin{equation}
{\bf D}=
\begin{pmatrix}
{\bf A}&{\bf B}\\
{\bf B}^{\dagger}&{\bf C}
\end{pmatrix}
,
\end{equation}
where ${\bf A}={\bf A}^{\dag},{\bf C}={\bf C}^{\dag}$ and ${\bf B}$ are $3\times3$ matrices. Following ref.~\cite{benatti}, it is possible to characterize the entanglement capabilities of the bath-mediated interaction between the two atoms of the cavity-QED system we are treating here. Namely, the condition to entangle the atomic subsystem using the Markovian dynamics described by Eq.~(\ref{master6}) is 
$\left({{\bf u}^*}{{\bf A}}{\bf u}\right)\left({\bf v^*}{{\bf C}^{T}}{\bf v}\right)<\modul{{\bf u^*}{{\cal R}e({\bf B})}{\bf v}}^2$,
where ${\bf u}=(\cos2\theta,-i,0)^T$ and ${\bf v}=(\cos2\varphi,i,0)^T$ carry information about the generic initial states of atoms $1$ and $2$, unitarily rotated by the angles $\theta$ and $\varphi$ around the $\bf z$ axis of the Bloch sphere.  For the case of Eq.~(\ref{master6}), we identify 
\begin{equation}
\label{benattimatrices}
{\bf A}={\bf C}=\gamma
\begin{pmatrix}
\frac{2N'+1}{4}&\frac{i}{4}&0\\
-\frac{i}{4}&\frac{2N'+1}{4}&0\\
0&0&0
\end{pmatrix}
\hskip0.2cm{{\bf B}}=\gamma
\begin{pmatrix}
\frac{M'}{2}&0&0\\
0&-\frac{M'}{2}&0\\
0&0&0
\end{pmatrix}
.
\end{equation}
While these matrices depend solely on the form of the reduced master equation for the atomic subsystem, the condition to entangle the two-level atoms depends on the initially prepared atomic state via the vectors ${\bf u}$ and ${\bf v}$, that carry information about $\theta,\,\varphi$. If the initial state of the atoms is taken to be $\ket{gg}_{12}\!\bra{gg}$, the condition becomes 
\begin{equation}
\label{mycondition}
N^2<M^2
\end{equation}
to entangle two atoms in cavities. Eq.~(\ref{mycondition}) has a clear physical interpretation: it is the sufficient and necessary condition for the entanglement of the driving field~\cite{simon}. If we start with the atoms in their ground states, the entanglement of the driving field represents, thus, the irremissible condition for the creation of entanglement between the atoms, despite the atomic spontaneous emission~\footnote{The condition we have found can be generalized to the case in which the local properties of the two-mode driving field are different. In particular, if the phase-insensitive parameters of modes $a$ and $b$ are $N_{a}\neq{N}_{b}$ (with the same value of the phase-sensitive $M$), then we get $N_{a}N_{b}<M^2$.}. To confirm the validity of this relation, we have solved the master equation~(\ref{master6}), looking for the set of Bloch equations for the matrix elements $\rho_{ijhk}=\sand{ij}{\rho_{12}}{hk}$ ($i,j,h,k=e,g$) under the assumption $M^2{<}N(N+1)$ and $\Gamma=0$, even though including $\Gamma\neq0$ is straightforward. Using the basis $\{\ket{ee},\ket{eg},\ket{ge},\ket{gg}\}_{12}$, we find the following set of coupled, linear differential equations
\begin{equation}
\label{blochequations}
\begin{aligned}
&{\dot\rho}_{eeee}=\gamma\left[-2n^{1}_{1}\rho_{eeee}+n^{0}_{1}(\rho_{egeg}+\rho_{gege})+2{M}\rho_{eegg}\right],\\
&{\dot\rho}_{egeg}=\gamma\left[n^{0}_{1}(1-\rho_{gege})+\rho_{eeee}-n^{1}_{3}\rho_{egeg}-2{M}\rho_{eegg}\right],\\
&{\dot\rho}_{gege}=\gamma\left[n^{0}_{1}(1-\rho_{egeg})+\rho_{eeee}-n^{1}_{3}\rho_{gege}-2{M}\rho_{eegg}\right],\\
&{\dot\rho}_{eegg}=-\gamma\left[n^{1}_{2}\rho_{eegg}-{M}(1-2\rho_{gege}-2\rho_{egeg})\right],
\end{aligned}
\end{equation}  
where, $n^{l}_{k}=(kN+l)$ and, by hermiticity, $\rho_{eegg}=\rho_{ggee}$.
All the other matrix elements are decoupled from these equations. 
The condition $Tr_{12}(\rho_{12})=1$ gives the equation for $\rho_{gggg}$. 

\begin{figure} [ht]
{\bf (a)}\hskip5cm{\bf (b)}
\centerline{\psfig{figure=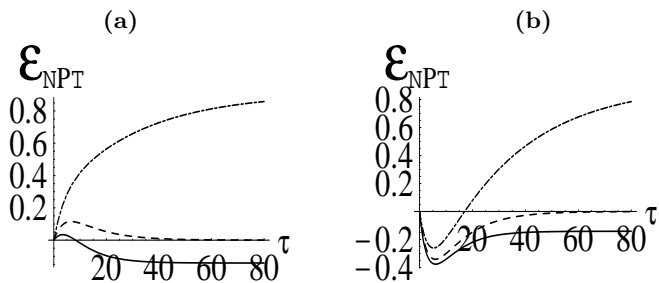,width=9.5cm,height=3.5cm}}
\caption{Entanglement of $\rho_{12}$ after the reduced Markovian dynamics 
described by the master equation~(\ref{master6}). The entanglement is plotted as a function of the dimensionless interaction time $\tau=\gamma{t}$ for $N=0.7$ and three different 
values of $M$: $M=0.79$ (solid curve), $M={\cal B}_{ss}(0.7)=0.902$ (dashed curve) and $M=\sqrt{N(N+1)}=1.09$ (dot-dashed curve). In {\bf (a)}, the initial state is $\ket{gg}_{12}\!\bra{gg}$ while in {\bf (b)}, it is $\ket{ee}_{12}\!\bra{ee}$.}
\label{benattipicture}
\end{figure}

Solving Eqs.~(\ref{blochequations}), we can find the dynamics of entanglement between atoms $1$ and $2$. We use the entanglement measure based on negativity of partial transposition (NPT), defined by
${\cal E}_{NPT}(t,N,M)=-2\lambda^{-}_{i}(t,N,M)$,
where $\lambda^{-}_{i}(t,N,M)$ is the negative eigenvalue of the partially transposed density matrix $\rho^{T_{2}}_{12}$ ($T_{2}$ indicates partial transposition with respect to $2$)~\cite{zyczkowski}. NPT is a necessary and sufficient condition for entanglement of any bipartite qubit system~\cite{Horodecki}. The results are shown in Fig.~\ref{benattipicture}.

By inspection of Fig.~\ref{benattipicture}{\bf (a)}, it is apparent that, even for $N<M$, the long time behavior of the entanglement function can lead to a separable atomic state. This is due to the fact that the condition found does not give information about the steady-state entanglement. The criterion for the entanglement due to the interaction with a Markovian bath is, indeed, relative to the creation of quantum correlation in an initially separable state, with a positive gradient of ${\cal E}_{NPT}(t,N,M)$. In order to be sure that the Markovian reduced dynamics will lead to the creation of entanglement, the trend of the entanglement function at the initial interaction time $t=0$ has to be positive. This is the case for $\rho_{12}(0)=\ket{gg}_{12}\!\bra{gg}$  but not for $\rho_{12}(0)=\ket{ee}_{12}\!\bra{ee}$, for example. For this initial state, it is $\partial_{t}{\cal E}_{NPT}(0,N,M)<0$, as can be seen in Fig.~\ref{benattipicture}{\bf (b)}, and the entangling condition leads to $N+1<M$, that is physically meaningless since $N+1>\sqrt{N(N+1)},\,\forall~{N}$. An analogous result is found for the asymmetric case in which one atom is prepared in $\ket{g}$ while the other is in $\ket{e}$. However, as stated before, this does not exclude the possibility that the atomic state becomes entangled later on. 

It is, thus, interesting to investigate the conditions that have to be fulfilled for the atomic steady state to be entangled. In particular, we are interested in finding the boundary value ${\cal B}_{ss}(N)$ of the phase-sensitive parameter $M$ beyond which we are sure that the atomic steady state is inseparable. To find ${\cal B}_{ss}(N)$ we have to look at the asymptotic behavior of the entanglement function that turns out to be a non-decreasing function of the parameter $M$, for a fixed value of $N$, as can be seen solving Eqs.~(\ref{blochequations}) and looking for the steady solutions. Thus, the condition 
\begin{equation}
\label{wonmincondition}
\begin{aligned}
&\lim_{t\rightarrow\infty}{\cal E}_{NPT}(t,N,M)\vert_{M={\cal B}_{ss}(N)}=0\\
\end{aligned}
\end{equation}
 fully characterizes the boundary value for $M$ that we are looking for. Using this, we find
\begin{equation}
\label{boundary}
{\cal B}_{ss}(N)=-\alpha+\sqrt{\alpha+N(N+1)}
\end{equation}
with $\alpha=1/4(\Delta{x})^2$ and $(\Delta{x})^2=N+1/2$ the variance of the in-phase quadrature of the bath. It is ${\cal B}_{ss}(N)\le\sqrt{N(N+1)},\,\forall\,N$. The boundary, Eq.~(\ref{boundary}), is independent from the initial preparation of the atoms. Thus, for the CQED system considered in this work, the atomic steady state is entangled when the two-mode driving field satisfies the condition ${\cal B}_{ss}(N)<M\le\sqrt{N(N+1)}$. Outside this range, the steady state is separable and the atomic subsystem may be entangled depending on the initial preparation of the atoms. This is one of the main results of this paper.

Eq.~(\ref{boundary}) represents a very strict condition on the properties of the driving field. The experimentally available source of squeezed light are, indeed, quite bright (that means large values of $N$) and it is easy to see that ${\cal B}_{ss}(N\gg1)\simeq\sqrt{N(N+1)}$, dramatically shrinking the range of values of $M$ in which the atomic steady state is entangled. 

\begin{figure} [t]
{\bf (a)}\hskip4cm{\bf (b)}
\centerline{\psfig{figure=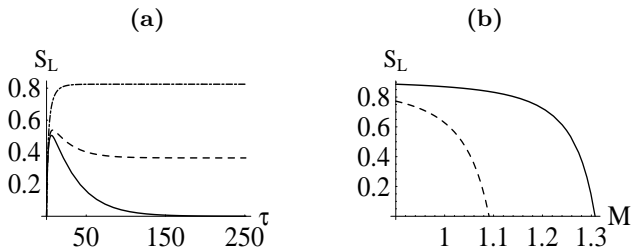,width=9.5cm,height=3.0cm}}
\caption{In {\bf (a)}, $S_{L}(t,N,M)$ is plotted against $\tau=\gamma{t}$ for $N=0.7$ and $M=\sqrt{N(N+1)}=1.09$ (solid line), $M=1.05$ (dashed line) and $M=0.8$ (dot-dashed line). In {\bf (b)}, $S_{L}(\infty,N,M)$ as a function of $M$ for $N=0.7$ (dashed line) and $N=0.9$ (solid line).}
\label{benattipicture2}
\end{figure}

Finally, it is worth stressing that the the range of values of $M$ in which the steady state is pure reduces to the point $M=\sqrt{N(N+1)}$. We look for the boundary value at which the atomic steady state becomes pure. As a measure for purity we take the {\it linearized entropy}, that for a two-qubit system is defined by $S_{L}(N,M,t)=4/3\left(1-Tr_{12}\left\{\rho^2_{12}\right\}\right)$ and ranges from 0 (pure states) to 1 (maximally mixed ones). Only the interaction with a pure squeezed reservoir realizes a pure atomic steady state. These results, for $\rho_{12}(0)=\ket{gg}_{12}\!\bra{gg}$, are shown in Fig.~\ref{benattipicture2}. In ${\bf (a)}$ $S_{L}$ is plotted versus the dimensionless interaction time for the case of a pure squeezed state (solid line), for $M=\sqrt{N(N+0.9)}$ (dashed line) and $M=\sqrt{N(N+0.2)}$ (dot-dashed line). In {\bf (b)}, we show the behavior of the linear entropy as a function of the phase-sensitive parameter $M$ for $N=0.7$ (dashed line) and $N=0.9$ (solid line). It is apparent that the higher is $N$, the more mixed is the atomic state, for a fixed $M$. This is because higher values of $N$ correspond to an increase of the thermal character of each field mode, in a two-mode squeezed state. On the other hand, increasing the squeezing properties of the driving ({\it i.e.}, increasing $M$) leads to a rapid decrease of the mixedness of the state. This is because the populations $\rho_{egeg}=\rho_{gege}$ decay if we increase $M$, a signature of the two-photon correlations in a highly squeezed two-mode state~\cite{massimoinfanzia}. 

{\it Remarks - }In this paper, we investigated about the mechanisms for the transfer of entanglement from a continuous-variable state to a qubit system. Starting from the recent proposal~\cite{kraus}, we have shown that the efficiency of the scheme depends on the cavity decay rate. We have addressed the conditions for the entanglement of an atomic subsystem induced by the Markovian interaction with a broad-band two-mode squeezed bath, both in the dynamical and steady conditions. We have found the sufficient and necessary conditions for the driving field to entangle two atoms in cavities. The entanglement of the driving field is only a necessary condition to entangle the two atoms, in steady state. We found the boundary value of $M$ beyond which the atomic steady state is entangled and pure. The effect of the atomic spontaneous emission rate $\Gamma$ can be included in our analysis and Eq.~(\ref{boundary}) is still valid replacing $N$ with $N'$. Our analysis gives some insight in the role played by purity and quantum correlation of the squeezed state in an entanglement-transfer process. 


{\it Acknowledgments - }We thank Dr. Z. Ficek for bringing his work to our attention. We acknowledge fruitful discussion with Prof. S. Swain and Dr. J. F. McCann. This work was in part supported by the UK Engineering and Physical Science Research Council grant GR/S14023/01 and the Korea Research Foundation  grant 2003-070-C00024. M.P. thanks the IRCEP (International Research Centre for Experimental Physics, Queen's University of Belfast) for supporting his studentship. 


\end{document}